\documentclass[12pt]{article}
\usepackage{amssymb,amsmath,epsfig}
\begin{document}

\title{\bf Cylinder with Charged Anisotropic Source}
\author{M. Sharif\thanks {msharif.math@pu.edu.pk}
and H. Ismat Fatima\thanks{hafizaismatfatima@yahoo.com}\\
Department of Mathematics, University of the Punjab,\\
Quaid-e-Azam Campus, Lahore-54590, Pakistan.}

\date{}

\maketitle
\begin{abstract}
We take charged anisotropic fluid cylinder when there is no
external pressure acting on the fluid. This is a cylindrical
version of the Krori and Barua's method to explore the field
equations with anisotropic fluid. We discuss models with positive
matter density and pressure that satisfy all the energy and
stability conditions. It is found that charge does not vanish at
the center of the cylinder. The equilibrium condition as well as
physical conditions are discussed. Further, we highlight the
connection between our solutions and the charged strange quark
stars as well as with dark matter including charged massive
particles. The graphical analysis of the matter variables versus
charge is given which indicates a physically reasonable matter
distribution.
\end{abstract}

\section{Introduction}

The study of general relativistic charged compact objects is of
fundamental importance in astrophysics. Strong magnetic fields,
different kinds of phase transitions and solid stellar core cause
anisotropy in the fluids. However, charged fluids with anisotropy
complicates the solution of the field equations. Equations of state
(EoS) has important consequences in such situations. Many exact
solutions have been obtained \cite{9} by using a simple form of the
energy-momentum tensor and assuming some symmetries.

There have been pervious discussions of a similar nature by Evan
\cite{9a}, Bronnikov \cite{9b}, Latelier and Tobensky \cite{9c},
and Kramer \cite{9d}. They used various EoS that could be written
in the form $\rho=\gamma p$ for specific positive values of
$\gamma$, as well as energy conservation. Some work has been done
on charged anisotropic static matter by using spherically
symmetric stars with the linear, nonlinear and Chaplygin gas EoS.
Ivanov \cite{1} showed that the field equations can be simplified
by using linear EoS for a charged perfect fluid but with
non-integrable equations. Sharma and Maharaj \cite{2} explored the
field equations for static spherically symmetric uncharged
anisotropic fluid with combined linear EoS and a particular mass
function.

Charged anisotropic fluids have been discussed in General
Relativity since the pioneering work of Bonner \cite{3}. Ray et
al. \cite {4} investigated charged anisotropic spheres with
Chaplygin gas EoS. Thirukkanesh and Maharaj \cite{5} generated
models for charged anisotropic spherically symmetric stars by
using linear EoS as well as choosing one of the metric functions
and electric field intensity. Horvat et al. \cite{6} studied
gravastars for charged anisotropic fluid. Recently, Victor et al.
\cite {7} explored solutions for the charged anisotropic spheres
with linear or nonlinear EoS.

Over the years, many authors have proposed various formulations to
solve the field equations for cylindrically symmetric spacetime.
Nilsson et al. \cite{8} investigated cylindrically symmetric perfect
fluid models. One of the authors (MS) \cite{10} explored perfect
fluid, static cylindrically symmetric solutions of the field
equations by using different EoS. Sharif and Fatima \cite{11} worked
for the charged anisotropic cylinder but they discussed
gravitational collapse. Som \cite{12} explored the charged dust
cylinder. However, there has been a little progress towards
investigating charged anisotropic static cylindrically symmetric
solutions with or without using an EoS.

In a recent paper \cite{12a}, we have explored exact solutions of
the field equations for the charged anisotropic static cylindrically
symmetric spacetime using Thirukkanesh and Maharaj \cite{5}
approach. Here we extend this study for the charged anisotropic
static cylindrically symmetric spacetime by using Victor et al.
\cite {7} procedure. A system of differential equations for matter
as well as electric field intensity and anisotropic pressures are
solved on the basis of linear and nonlinear EoS. Numerical factors
depending on matching conditions are used with each EoS which
provide relationship among charge distribution, pressure anisotropy
and EoS.

The outline of the paper is as follows: In the next section, we
write down the Einstein-Maxwell field equations for the static
cylindrically symmetric spacetime and also express this system of
equations with original Krori and Barua's \cite{13} assumptions.
We apply the central and boundary conditions on electric field
intensity and radial pressure respectively to analyze these field
equations at center and on the boundary of the cylinder. Section
\textbf{3} investigates models for linear, nonlinear and Chaplygin
gas EoS by taking positive matter densities and pressures
corresponding to the relevant EoS. In section \textbf{4}, we match
smoothly the interior and exterior metrics and bring
adimesionality in the three models. Section \textbf{5} provides
some physical features of these models. In particular, we discuss
the stability conditions, energy conditions, the ven der Waals EoS
\cite{14} and the equilibrium conditions for our models. The last
section \textbf{6} contains concluding remarks about the results.

\section{The Field Equations}

We take the static cylindrically symmetric spacetime given by
\cite{15}
\begin{equation}\label{1}
ds^{2}=e^{2\nu}dt^{2}-e^{2\mu-2\nu}dr^{2}-r^{2}e^{-2\nu}d\phi^{2}-e^{2\mu-2\nu}dz^{2},
\end{equation}
where $\nu$ and $\mu$ are functions of $r$. The transformation
$d\phi=e^{\nu}d\theta$ leads the above equation to the following
form \cite{8, 11}
\begin{equation}\label{2}
ds^{2}=e^{2\nu}dt^{2}-e^{2\mu-2\nu}dr^{2}-r^{2}d\theta^{2}-e^{2\mu-2\nu}dz^{2}.
\end{equation}
The field equations for the charged anisotropic source are
\begin{equation}\label{3}
R_{ab}-\frac{1}{2}g_{ab}R=\kappa(T^{(m)}_{ab}+T^{(em)}_{ab}),
\end{equation}
where $T^{(m)}_{ab}$ and $T^{(em)}_{ab}$ are the energy-momentum
tensors for anisotropic matter and electromagnetic field
respectively. The energy-momentum tensor for anisotropic fluid is
\begin{equation}\label{4}
T^{(m)}_{ab}=(\rho+p_{r})u_{a}u_{b}-p_{t}g_{ab}+(p_{t}-p_{r})\eta_{a}
\eta_{b}
\end{equation}
satisfying $u^{a}u_{a}=-\eta^{a} \eta_{a}=1$, where $\rho$ is the
charge density, $p_{r}$ is the radial pressure, $p_{t}$ is the
tangential pressure, $u_{a}=e^{\nu}\delta^{0}_{a}$ is the
4-velocity and $\eta _{a}=-e^{\mu-\nu}\delta_{a}^{1}$ is the
4-unit vector. The energy-momentum tensor for the electromagnetic
field is
\begin{equation}\label{5}
T^{(em)}_{ab}=\frac{1}{4\pi}(-g^{cd}F_{ac}F_{bd}+\frac{1}{4}g_{ab}F_{cd}F^{cd}),
\end{equation}
where $F_{ab}=A_{b,a}-A_{a,b}$ is the Maxwell field tensor and $A_a$
is the 4-potential. The Maxwell field equations are given by
\begin{equation}\label{6}
[\sqrt{-g}F^{ab}]_{,b}=4\pi J^{a}\sqrt{-g},\quad F_{[ab,c]}=0,
\end{equation}
where $J^{a}=\sigma u^{a}$ is the 4-current of the fluid element and
$\sigma$ is the proper charge density.

The field equations for the line element (\ref{2}) become
\begin{eqnarray}\label{7}
e^{2\nu-2\mu}(\nu''-\mu'')=8\pi\rho+E^{2},\\\label{8}
\frac{e^{2\nu-2\mu}}{r}(-r\nu'^{2}+r\nu'\mu'+\mu')=8\pi
p_{r}-E^{2},\\\label{9} e^{2\nu-2\mu}(\nu'~^{2}+\mu'') =8\pi
p_{t}+E^{2},\\\label{10} \sigma=\frac{e^{2\nu-2\mu}}{4\pi
r}(re^{\mu-\nu}E)',
\end{eqnarray}
where prime denotes differentiation with respect to $r$ and
$E=2\sqrt{\pi}e^{-\mu}\frac{\partial A}{\partial r}$ stands for em
part. In the system of equations (\ref{7})-(\ref{10}), there are
seven unknowns, so we make physically reasonable choices for any two
of the unknowns. We take the gravitational potential $e^{2\mu-2\nu}$
and the electric field intensity $E$ as \cite{6}
\begin{eqnarray}\label{11}
e^{2\mu-2\nu}&=&\frac{1+(c_{1}-c_{2})r}{1+c_{1}r}, \\\label{12}
E^{2}&=&\frac{k(3+c_{1}r)}{(1+c_{1}r)^{2}},
\end{eqnarray}
where $c_1,~c_2$ and $k$ are constants. Substituting Eqs.(\ref{11})
and (\ref{12}) in Eq.(\ref{10}), we obtain
\begin{equation}\label{13}
\sigma\approx\frac{\sqrt{3k}}{2\pi r}.
\end{equation}
This shows that there is a singularity in the charge distribution at
$r=0$. However, this choice keeps the charge distribution regular at
the centre of the cylinder as $E(r)$ remains finite there
(\ref{12}).

The singularity free models for charged anisotropic static cylinder
are constructed by taking \cite{13}
\begin{eqnarray}\label{14}
\mu=Ar^{2},\quad \nu=Br^{2}+C,
\end{eqnarray}
where $A,~B$ and $C$ are constants. Using these values in
Eqs.(\ref{7})-(\ref{10}), it follows that
\begin{eqnarray}\label{15}
e^{2r^{2}(B-A)+2C}(2B-2A)=8\pi \rho +E^{2},\\\label{16}
e^{2r^{2}(B-A)+2C}(-4B^{2}r^{2}+4ABr^{2}+2A)=8\pi
p_{r}-E^{2},\\\label{17} e^{2r^{2}(B-A)+2C}(4B^{2}r^{2}+2A)=8\pi
p_{t}+E^{2},\\\label{18} \sigma=\frac{e^{2r^{2}(B-A)+2C}}{4\pi
r}(re^{r^{2}(A-B)-C}E)'.
\end{eqnarray}
Now we impose the central and boundary conditions on $E(r)$ and
$p_{r}(r)$ respectively as follows:
\begin{eqnarray}\label{19}
E(0)=0, \quad p_{r}(a)=0,
\end{eqnarray}
where $a$ is a positive constant and $r=a$ is the interface of the
charged fluid and vacuum (i.e., boundary of the cylinder). We
apply central conditions to the system of
Eqs.(\ref{15})-(\ref{17}), it follows that
\begin{eqnarray}\label{20}
\rho (0)=\frac{(B-A)e^{2C}}{4\pi},\quad
p_{r}(0)=\frac{Ae^{2C}}{4\pi},\quad p_{t}(0)=\frac{Ae^{2C}}{4\pi}.
\end{eqnarray}
This shows that $p_{r}(r)=p_{t}(r)$ at $r=0$, hence the anisotropy
of the cylinder vanishes at the center. Applying the boundary
conditions to Eqs.(\ref{15})-(\ref{17}), we get
\begin{eqnarray}\label{23}
\rho
(a)&=&\frac{e^{2a^{2}(B-A)+2C}(-2B^{2}a^{2}+B+2ABa^{2})}{4\pi},\\\label{24}
p_{t}(a)&=&\frac{e^{2a^{2}(B-A)+2C}(A+ABa^{2})}{2\pi},\\\label{25}
E^{2}(a)&=&e^{2a^{2}(B-A)+2C}(4ABa^{2}+2A-4B^{2}a^{2}).
\end{eqnarray}
The general expressions for $p_{t}$ and $E^{2}$ from
Eqs.(\ref{15})-(\ref{17}) are
\begin{eqnarray}\label{26}
p_{t}(r)&=&\frac{e^{2r^{2}(B-A)+2C}(-2B+2B^{2}r^{2}+2A)}{4\pi}+\rho
,\\\label{27} E^{2}(r)&=&e^{2r^{2}(B-A)+2C}(2B-2A)-8\pi \rho.
\end{eqnarray}

\section{Models for Equations of State}

The general form of EoS is
\begin{equation}\label{28}
p_{r}=p_{r}(\rho ,a_{1},a_{2}),
\end{equation}
where $a_{1}$ and $a_{2}$ are parameters constrained by
\begin{eqnarray}\label{29}
p_{r}(0)=p_{r}[\rho(0),a_{1},a_{2}],\quad
0=p_{r}[\rho(a),a_{1},a_{2}].
\end{eqnarray}
Adding Eqs.(\ref{15}) and (\ref{16}), we obtain
\begin{equation}\label{30}
\rho+p_{r}=\frac{e^{2r^{2}(B-A)+2C}}{8\pi}(-4B^{2}r^{2}+2B+4ABr^{2})\equiv
l(r).
\end{equation}
This equation may be used with the assumed EoS to find $\rho$ and
$p_{r}$. The corresponding value of $\rho$ will be used in
Eqs.(\ref{26}) and (\ref{27}) to evaluate $p_{t}$ and $E^{2}$
respectively. In the following we discuss three types of EoS:

\subsection{The Linear EoS}

This is given by
\begin{equation}\label{31}
p_{r}=\alpha_{1}+\alpha_{2}\rho,
\end{equation}
where $\alpha_1,~\alpha_2$ are constants. Using this EoS, we obtain
expressions for $\rho,~p_{r},~p_{t}$ and $E^{2}$ as follows:
\begin{eqnarray}\label{32}
\rho&=&\frac{e^{2r^{2}(B-A)+2C}(-4B^{2}r^{2}+2B+4ABr^{2})-8\pi
\alpha_{1} }{8\pi(1+\alpha_{2})},\\\label{33}
p_{r}&=&\frac{\alpha_{2}e^{2r^{2}(B-A)+2C}(-4B^{2}r^{2}+2B+4ABr^{2})+8\pi
\alpha_{1}}{1+\alpha_{2}}.
\end{eqnarray}
Using Eq.(\ref{32}) in Eqs.(\ref{26}) and (\ref{27}) successively,
we get
\begin{eqnarray}\label{34}
p_{t}&=&\frac{e^{2r^{2}(B-A)+2C}((2B-4A-4B^{2}r^{2})(1+\alpha_{2})
+4B^{2}r^{2}-4ABr^{2}-2B)}{8\pi
(1+\alpha_{2})}\nonumber\\&-&\frac{\alpha_{1}}{1+\alpha_{2}},
\\\label{35}
E^{2}&=&\frac{e^{2r^{2}(B-A)+2C}((2B-2A)(1+\alpha_{2})+4B^{2}r^{2}-4ABr^{2}-2B)}{
1+\alpha_{2}}-\frac{8\pi \alpha_{1}}{1+\alpha_{2}}.\nonumber\\
\end{eqnarray}
The values of constants $\alpha_{1}$ and $\alpha_{2}$ are found by
solving Eqs.(\ref{29}) and (\ref{31}) as
\begin{equation}\label{36}
\alpha_{1}=-\frac{\rho(a)p_{r}(0)}{\rho(0)-\rho(a)},\quad
\alpha_{2}=\frac{p_{r}(0)}{\rho(0)-\rho(a)}.
\end{equation}

\subsection{The Nonlinear EoS}

The nonlinear EoS is given by
\begin{eqnarray}\label{37}
p_{r}=\beta_{1}+\frac{\beta_{2}}{\rho^{n}},
\end{eqnarray}
where $n\neq-1$ and $\beta_1,~\beta_2$ are constants. It is a
modification of the Chaplygin gas EoS used by Bertolami and Paramos
\cite{16} to describe neutral dark stars. For $n=1$, Eqs.(\ref{30})
and (\ref{37}) lead to
\begin{equation}\label{38}
\rho=\frac{l(r)-\beta_{1}\pm
\sqrt{(l(r)-\beta_{1})^{2}-4\beta_{2}}}{2}.
\end{equation}
Substituting Eq.(\ref{38}) in Eqs.(\ref{37}), (\ref{26}) and
(\ref{27}) respectively, it follows that
\begin{eqnarray}\label{39}
p_{r}&=&\beta_{1}+\frac{2\beta_{2}}{l(r)-\beta_{1}\pm
\sqrt{(l(r)-\beta_{1})^{2}-4\beta_{2}}}, \\\label{40}
p_{t}&=&\frac{e^{2r^{2}(B-A)+2C}(-B+2B^{2}r^{2}+2A)}{4\pi}
\nonumber\\&+&\frac{l(r)-\beta_{1}\pm
\sqrt{(l(r)-\beta_{1})^{2}-4\beta_{2}}}{2} ,\\\label{41}
E^{2}(r)&=&e^{2r^{2}(B-A)+2C}(2B-2A)-4\pi
\left(l(r)-\beta_{1}\right.\nonumber\\
&\pm&\left.\sqrt{(l(r)-\beta_{1})^{2}-4\beta_{2}}\right),
\end{eqnarray}
where $l(r)$ is given by Eq.(\ref{30}). The constants $\beta_{1}$
and $\beta_{2}$ are found from Eqs.(\ref{37}) and (\ref{29}) as
\begin{equation}\label{42}
\beta_{1}=\frac{\rho(0)p_{r}(0)}{\rho(0)-\rho(a)},\quad
\beta_{2}=-\frac{\rho(0)\rho(a)p_{r}(0)}{\rho(0)-\rho(a)}.
\end{equation}
Equation (\ref{38}) implies that $\beta_{2}$ must be negative so
that each root in this equation has definite sign which will
correspond to a positive definite matter density.

\subsection{The Modified Chayplygin Gas EoS}

This EOS has the following form
\begin{equation}\label{43}
p_{r}=\gamma_{1}\rho+\frac{\gamma_{2}}{\rho},
\end{equation}
where $\gamma_1,~\gamma_2$ are constants. This is used to describe
static, neutral, phantom-like sources \cite{17}. Using this EoS
with Eq.(\ref{30}), we get
\begin{equation}\label{44}
\rho=\frac{l(r)\pm
\sqrt{l(r)^{2}-4(1+\gamma_{1})\gamma_{2}}}{2(1+\gamma_{1})}.
\end{equation}
Substituting this value of $\rho$ in Eq.(\ref{43}) as well as in
Eqs.(\ref{26}) and (\ref{27}) successively, it follows that
\begin{eqnarray}\label{45}
p_{r}&= &\gamma_{1}\left(\frac{l(r)\pm
\sqrt{l(r)^{2}-4(1+\gamma_{1})\gamma_{2}}}{2(1+\gamma_{1})}\right)
\nonumber\\&+&\frac{2\gamma_{2}(1+\gamma_{1})} {l(r)\pm
\sqrt{l(r)^{2}-4(1+\gamma_{1})\gamma_{2}}},\\\label{46}
p_{t}&=&\frac{e^{2r^{2}(B-A)+2C}(-B+2B^{2}r^{2}+2A)}{4\pi}\nonumber\\&+&\frac{l(r)\pm
\sqrt{l(r)^{2}-4(1+\gamma_{1})\gamma_{2}}}{2(1+\gamma_{1})}
,\\\label{47} E^{2}&=&e^{2r^{2}(B-A)+2C}(2B-2A)\nonumber\\&-&4\pi
\left(\frac{l(r)\pm
\sqrt{l(r)^{2}-4(1+\gamma_{1})\gamma_{2}}}{1+\gamma_{1}}\right),
\end{eqnarray}
where $\gamma_{1}$ and $\gamma_{2}$ are
\begin{equation}\label{48}
\gamma_{1}=\frac{\rho(0)p_{r}(0)}{\rho(0)^{2}-\rho(a)^{2}}, \quad
\gamma_{2}=-\frac{\rho(0)p_{r}(0)\rho(a)^{2}}{\rho(0)^{2}-\rho(a)^{2}}.
\end{equation}
For $(1+\gamma_{1})<0$ and $\gamma_{2}>0$ or $(1+\gamma_{1})>0$ and
$\gamma_{2}<0$, Eq.(\ref{44}) yields roots of definite sign which
will again correspond to the positive matter densities. It is clear
that the positive or negative matter densities depend upon the
positivity or negativity of the constants
$\beta_{1},~\beta_{2},~\gamma_{1}$ and $\gamma_{2}$, i.e., the roots
of Eqs.(\ref{38}) and (\ref{44}) respectively.

\section{Matching Conditions and Adimensional Matter Sources}

Here we take the charged static cylindrically symmetric spacetime as
an exterior region given by \cite{18}
\begin{equation}\label{49}
ds^{2}=N(r)dt^{2}-\frac{1}{N(r)}dr^{2}-r^{2}d\theta^{2}-r^{2}d\psi^{2},\quad
N(r)=\frac{q^{2}}{r^{2}}-\frac{2m}{r},
\end{equation}
where $q$ and $m$ are charge and mass respectively. Using the
transformation, $d\psi=\frac{1}{\sqrt{N(r)}}d\phi$, this takes the
form
\begin{equation}\label{50}
ds^{2}=N(r)dt^{2}-\frac{1}{N(r)}dr^{2}-r^{2}d\theta^{2}-\frac{r^{2}}{N(r)}d\phi^{2}.
\end{equation}
With the radial transformation $r=\frac{m^{2}-q^{2}}{r'}$, it
becomes
\begin{equation}\label{51}
ds^{2}=\frac{\frac{q^{2}}{r'^{2}}-\frac{2m^{3}}{r'^{3}}+\frac{2mq^{2}}{r'^{3}}}
{(\frac{m^{2}-q^{2}}{r'^{2}})^{2}}dt^{2}-\frac{(\frac{m^{2}-q^{2}}{r'^{2}})^{4}}
{\frac{q^{2}}{r'^{2}}
-\frac{2m^{3}}{r'^{3}}+\frac{2mq^{2}}{r'^{3}}}(dr'^{2}+d\phi^{2})-r'^{2}d\theta^{2}.
\end{equation}

To match the interior metric (\ref{2}) with the exterior
(\ref{51}), we impose the continuity of $g_{00},~g_{11}$ and
$\frac{\partial g_{00}}{\partial r}$ across a surface at $r'=a$ by
using the procedure \cite{19}. In our case, this yields the
following expressions for $A,~B$ and $C$ in terms of adimensional
parameters $\eta=\frac{m}{a}$ and $\chi=\frac{|q|}{a}$ as
\begin{eqnarray}\label{52}
A&=&\frac{\ln(\eta^{2}-\chi^{2})}{a^{2}(\eta^{2}-\chi^{2})^{2}},\\\label{53}
B&=&-\frac{1}{a^{2}(\eta^{2}-\chi^{2})}(\frac{1}{2}
+\frac{\chi^{2}}{4(\chi^{2}-2\eta^{3}+\eta\chi^{2})}),\\\label{54}
C&=&\frac{1}{2}
+\frac{\chi^{2}}{4(\chi^{2}-2\eta^{3}+\eta\chi^{2})}
+\ln\frac{\eta^{2}-\chi^{2}}{\sqrt{\chi^{2}-2\eta^{3}+\eta\chi^{2}}}.
\end{eqnarray}
We can see from Eq.(\ref{14}) that $A$ and $B$ have dimension of
$length^{-2}$ and $C$ is dimensionless. It is very important that
the field equations can eventually be expressed in terms of these
adimensional constants and the dimensionless radial coordinate
$x=\frac{r}{a}$. Here adimesionality is denoted by $hats$. We
assume that the interior of the fluid cylinder is described by
$x\in [0,1)$. We reformulate all models as adimensional models
where we are denoting
$\hat{A}=a^{2}A,~\hat{B}=a^{2}B,~\hat{\rho}=a^{2}\rho,~\hat{p}_{r}=a^{2}p_{r},~
\hat{p}_{t}=a^{2}p_{t},~\hat{E}^{2}=a^{2}E^{2}$ and
$\hat{\sigma}=a^{2}\sigma$. The quantities which are originally
dimensionless are denoted by the actual symbol. The central and
the boundary conditions at $x\in [0,1)$ become
\begin{eqnarray}\label{55}
\hat{\rho} (0)&=&\frac{(\hat{B}-\hat{A})e^{2C}}{4\pi},\\\label{56}
\hat{p_{r}}(0)&=&\frac{\hat{A}e^{2C}}{4\pi},\\\label{57}
\hat{p_{t}}(0)&=&\frac{\hat{A}e^{2C}}{4\pi},\\\label{58}
\hat{\rho} (1)&=&\frac{e^{2(\hat{B}-\hat{A})+2C}(2\hat{B}^{2}
-\hat{B}-2\hat{A}\hat{B})}{4\pi},\\\label{59}
\hat{p}_{t}(1)&=&\frac{e^{2(\hat{B}-\hat{A})+2C}(\hat{A}+\hat{A}\hat{B})}{2\pi},\\\label{60}
\hat{E}^{2}(1)&=&e^{2(\hat{B}-\hat{A})+2C}(4\hat{A}\hat{B}+2\hat{A}-4\hat{B}^{2}).
\end{eqnarray}

\subsubsection*{The Adimensional linear EoS model}

The adimensional linear EoS is given by
\begin{eqnarray}\label{61}
\hat{p}_{r}&=&\hat{\alpha}_{1}+\alpha_{2}\hat{\rho}.
\end{eqnarray}
The corresponding quantities will become
\begin{eqnarray}\label{62}
\hat{\rho}&=&\frac{e^{2x^{2}(\hat{B}-\hat{A})+2C}(-4\hat{B}^{2}x^{2}+2\hat{B}
+4\hat{A}\hat{B}x^{2})-8\pi\hat{\alpha}_{1}}{8\pi
(1+\alpha_{2})},\\\label{63}
\hat{p}_{r}&=&\frac{8\pi\hat{\alpha}_{1}
+\alpha_{2}e^{2x^{2}(\hat{B}-\hat{A})+2C}(-4\hat{B}^{2}x^{2}+2\hat{B}
+4\hat{A}\hat{B}x^{2})}{1+\alpha_{2}},\\\label{64}
\hat{p}_{t}&=&\frac{e^{2x^{2}(\hat{B}-\hat{A})+2C}}{8\pi}
(-2\hat{B}+4\hat{B}^{2}x^{2}+4\hat{A})+\hat{\rho},\\\label{65}
\hat{E}^{2}&=&e^{2x^{2}(\hat{B}-\hat{A})+2C}(2\hat{B}-2\hat{A})-8\pi
\hat{\rho},
\end{eqnarray}
where
\begin{eqnarray}\label{66}
\hat{\alpha}_{1}=-\frac{\hat{\rho}(1)\hat{p}(0)}{\hat{\rho}(0)
-\hat{\rho}(1)},\quad
\alpha_{2}=\frac{\hat{p}_{r}(0)}{\hat{\rho}(0) -\hat{\rho}(1)}.
\end{eqnarray}

\subsubsection*{The Adimensional Nonlinear EoS Model}

Here we have
\begin{eqnarray}\label{67}
\hat{p}_{r}&=&\hat{\beta}_{1}+\frac{\hat{\beta}_{2}}{\hat{\rho}}.
\end{eqnarray}
For this model, the corresponding quantities take the form
\begin{eqnarray}\label{68}
\hat{\rho}&=&\frac{\hat{l}(x)-\hat{\beta}_{1}\pm
\sqrt{(\hat{l}(x)-\hat{\beta}_{1})^{2}-4\hat{\beta}_{2}}}{2},\\\label{69}
\hat{p}_{t}&=&\frac{e^{2x^{2}(\hat{B}-\hat{A})+2C}(-\hat{B}
+2\hat{B}^{2}x^{2}+2\hat{A})}{4\pi}+ \hat{\rho} ,\\\label{70}
\hat{E}^{2}&=&e^{2x^{2}(\hat{B}-\hat{A})+2C}(2\hat{B}-2\hat{A})-8\pi
\hat{\rho},
\end{eqnarray}
where $\hat{\beta}_{1}$ and $\hat{\beta}_{2}$ are
\begin{eqnarray}\label{71}
\hat{\beta}_{1}=\frac{\hat{\rho}(0)\hat{p}_{r}(0)}{\hat{\rho}(0)-\hat{\rho}(1)},
\quad
\hat{\beta}_{2}=-\frac{\hat{\rho}(0)\hat{\rho}(1)\hat{p}_{r}(0)}{\hat{\rho}(0)-\hat{\rho}(1)}.
\end{eqnarray}

\subsubsection*{The Adimensional Modified Chaplygin EoS Model}

The adimensional modified Chaplygin EoS model
\begin{eqnarray}\label{72}
\hat{p}_{r}&=&\gamma_{1}\hat{\rho}+\frac{\hat{\gamma}_{2}}{\hat{\rho}}
\end{eqnarray}
yield the following quantities
\begin{eqnarray}\label{73}
\hat{\rho}&=&\frac{\hat{l}(x)\pm
\sqrt{\hat{l}(x)^{2}-4(1+\gamma_{1})\hat{\gamma}_{2}}}{2(1+\gamma_{1})},\\\label{74}
\hat{p}_{t}&=&\frac{e^{2x^{2}(\hat{B}-\hat{A})+2C}(-\hat{B}+2\hat{B}^{2}x^{2}
+2\hat{A})}{4\pi}+\hat{ \rho} ,\\\label{75}
\hat{E}^{2}&=&e^{2x^{2}(\hat{B}-\hat{A})+2C}(2\hat{B}-2\hat{A})-8\pi
\hat{\rho},
\end{eqnarray}
where
\begin{equation}\label{76}
\gamma_{1}=\frac{\hat{\rho}(0)\hat{p}_{r}(0)}{\hat{\rho}^{2}(0)-\hat{\rho}^{2}(1)},\quad
\hat{\gamma}_{2}=-\frac{\hat{\rho}(0)\hat{p}_{r}(0)\hat{\rho}^{2}(1)}
{\hat{\rho}^{2}(0)-\hat{\rho}^{2}(1)}.
\end{equation}
The adimensional proper charge density for all the above three
models will become
\begin{equation}\label{77}
\hat{\sigma}=\frac{e^{2x^{2}(\hat{B}-\hat{A})+2C}}{2\pi
x}\frac{d}{dx}(xe^{x^{2}(\hat{B}-\hat{A})+2C}\hat{E}).
\end{equation}

\section{Some Features of the Models}

In this section, we discuss some insights of the three models. The
exterior metric (\ref{49}) implies that singularity occurs at
$r=0,~\frac{q^2}{2m}$. It is regular everywhere except at $r=0$
\cite{18}. The surface with $r=\frac{q^{2}}{2m}$ describes a right
singular circular cylinder. Using Eqs.(\ref{52})-(\ref{54}) in
$\rho(0)$ of Eq.(\ref{20}), it turns out to be positive. Following
\cite{7}, we have $\eta=\frac{GM}{c^{2}a}\approx 1.147$ for $M$ to
be the solar mass and $a=1.48km$. For this value of $\eta$,
$\hat{\rho}(0),~\hat{\rho}(1),~\hat{p}_{t}(0),~\hat{p}_{t}(1),~\hat{E}(1)$
and $\hat{p}_{r}(0)$ turn out to be functions of $\chi$ only.
Similarly, the expressions for
$\hat{\alpha}_{1},~\alpha_{2},~\hat{\beta}_{1},~\hat{\beta}_{2},~\gamma_{1}$
and $\hat{\gamma}_{2}$ (EoS parameters in adimensional version) also
depend only on $\chi$. The analysis of $A,~B$ and $C$ implies that
the values of $\chi$ are restricted by the values of $\eta$ such
that $\chi<\eta$ for $\eta>0$.

Figures \textbf{1} and \textbf{2} indicate that the central density
$\hat{\rho}(0)$ and central pressure $\hat{p}_{r}(0)$ are
monotonically decreasing with increasing values of $\chi$ for
$\eta=1.147$ and $\chi\in(0,0.7]$. We see that $\hat{p}_{r}(0)>0$
only when $\chi\in[0,0.57)$, hence we take the maximum charge
$\chi=0.56$ to discuss our models. Figures \textbf{3-5} show that
$\hat{\beta}_{2}<0,~ 1+\hat{\gamma}_{1}>0$ and $\hat{\gamma}_{2}<0$
for the same value of $\eta$. We can have positive definite roots
from Eqs.(\ref{68}) and (\ref{73}), hence we can analyse the models
with positive matter density $\hat{\rho}(x)$. It is obvious that
$\chi=0$ implies no charge. If we increase $\chi$, it increases
repulsive electrostatic forces and consequently pressure and density
change.

Now we explore the behavior of matter sources for our models at
different charges. Firstly, we investigate the sources with linear
EoS. Figures \textbf{6-9} show the corresponding matter density,
radial pressure, tangential pressure and electric field intensity.
The graphs of density and pressure show the increasing behavior
while the electric field intensity is decreasing with the increasing
values of $\chi$. Further, $\hat{E}(x)$ decreases at every point in
the interval $x\in(0,0.1]$ with increasing $\chi$. Figure
\textbf{10} shows the behavior of charge density $\hat{\sigma}$
which is unbounded for each value of $\chi$.

The analysis for models with nonlinear and Chaplygin gas EoS
indicates that these models are similar to the model satisfying
linear EoS. These models correspond to the decreasing matter
densities and pressures. Each EoS affects the dependance of the
measure of anisotropy $\hat{\delta}$ on $x$. The only difference
arising from the three models is about the measure of anisotropy
$\hat{\delta}=\hat{p}_{t}-\hat{p}_{r}$. Figure \textbf{11} displays
the anisotropic parameter for the model corresponding to the linear
EoS which is increasing with the increasing $\chi$. Figures
\textbf{12} and \textbf{13} show the anisotropic parameters
corresponding to the nonlinear and Chaplygin gas EoS respectively.
From figures, we see that the anisotropic parameter for nonlinear
EoS is increasing while for the Chaplygin gas EoS, it is decreasing
with increasing $\chi$. In addition, Eq.(\ref{18}) contains proper
charge density $\sigma$ which implies that
\begin{eqnarray}\label{79}
E(r)=\frac{4\pi re^{\nu-\mu}}{r^{2}}\int_{0}^{r}r\sigma
e^{2\mu-2\nu}dr=\frac{q(r)}{r^{2}},
\end{eqnarray}
where
\begin{equation}\label{80}
q(r)=4\pi r e^{\nu-\mu}\int_{0}^{r}r\sigma e^{2\mu-2\nu}dr
\end{equation}
which is the net charge inside the cylinder of radius $r$.

For our cylindrically symmetric fluid models,
Eqs.(\ref{7})-(\ref{10}) represent the basic source parameters. We
formulate table \textbf{I} by computing adimensional values for
these basic sources. Approximate numerical values for central
density $\hat{\rho}(0)$, central pressure $\hat{p}(0)$, tangential
pressure $\hat{p}_{t}(1)$ and electric field intensity
$\hat{E}^{2}(1)$, are shown in table \textbf{I}. We note that the
maximum value of $\hat{\rho}(0)$ and the value of $\hat{p}_{r}(0)$
are smaller than maximum of $\hat{p}_{t}(1)$. These maxima
correspond to zero net charge sources. The source variables
$\hat{\rho}(0),~\hat{p}_{r}(0)$ with maximum charge ($\chi=0.56$)
have been changed while $\hat{p}_{t}(0)$ changes its sign at the
maximum charge. To compensate the stronger electric repulsion, this
sign inversion of tangential pressure is necessary. This is summarized
in the following table:\\\\
\textbf{Table I.} Approximate numerical values of some quantities.
\begin{center}
\begin{tabular}{|c|c|c|c|c|c|c|c|c|c|c|}
\hline   \textbf{$\chi$} & \textbf{$\hat{\rho}(0)$} &
\textbf{$\hat{p}(0)$} & \textbf{$\hat{p}_{t}(1)$}  &
\textbf{$\hat{E}^{2}(1)$}\\\hline 0 & 0.0221364 & 0.00651431&
0.00830257 & 0.266397\\\hline
0.56& 0.0172785 & 0.0000738598 & -0.0000726 & 0.370451\\
\hline
\end{tabular}
\end{center}

Now we discuss some consequences of our results.

\subsection{Stability}

Bertolami and Paramos \cite{16} argued that if the generalized
Chaplygin gas tends to a smooth distribution over space then most
density perturbations tend to be flattened within a time scale
related to their initial size and the characteristic speed of sound.
One of the important "physical acceptability conditions" for
anisotropic matter is that the squares of radial and tangential
sound speeds ($u_{r}^{2}=\frac{dp_{r}}{d\rho}$ and
$u_{t}^{2}=\frac{dp_{t}}{d\rho}$) should be less than the speed of
light \cite{20}. We explore it for the linear EoS model. The graphs
of the squares of radial and tangential sound velocities are shown
in Figures \textbf{14} and \textbf{15} respectively. These indicate
that $u_{r}^{2}$ is independent of $x$ and decreases with increasing
$\chi$ while $u_{t}^{2}$ monotonically increases with increasing $x$
and also increases with increasing $\chi$ for fixed $x$. For three
particular values considered here, these parameters satisfy
$0<u_{r}^{2}<1$ and $0<u_{t}^{2}<1$ everywhere within the charged
fluid.

Now we use Herrera \cite{20} and Andreasson's \cite{21} approach
to identify potentially unstable or stable anisotropic matter
configuration. According to their approach,
$|u_{t}^{2}-u_{r}^{2}|\leq1$  as shown in Figure \textbf{16}. This
implies that
\begin{equation*}
(i)\quad-1\leq u_{t}^{2}-u_{r}^{2}\leq0,\quad (ii)\quad
0<u_{t}^{2}-u_{r}^{2}\leq 1.
\end{equation*}
The first expression corresponds to the potentially stable model
which is obvious from Figure \textbf{17}, while the second
expression corresponds to the potentially unstable model. We note
that our model satisfy the potentially stable condition. If the
graph of $u_{t}^{2}-u_{r}^{2}$ keeps the same sign everywhere within
a matter distribution, there will be no cracking and the system is
stable. If there is a change of sign then it is alternating
potentially unstable to stable region within the matter distribution
and vice versa.

\subsection{Energy Conditions}

The energy conditions of the charged anisotropic fluid, the weak
energy condition, the strong energy condition and the dominant
energy condition are satisfied if and only if the following
inequalities hold:
\begin{eqnarray}\label{81}
\hat{\rho}+\hat{p}_{r}\geq 0,\\\label{82}
\hat{\rho}+\frac{\hat{E}^{2}}{8\pi}\geq 0,\\\label{83}
\hat{\rho}+\hat{p}_{t}+\frac{\hat{E}^{2}}{4\pi}\geq 0,\\\label{84}
\hat{\rho}+\hat{p}_{r}+2\hat{p}_{t}+\frac{\hat{E}^{2}}{4\pi}\geq
0,\\\label{85}
\hat{\rho}+\frac{\hat{E}^{2}}{8\pi}-|\hat{p}_{r}-\frac{\hat{E}^{2}}{8\pi}|\geq
0,\\\label{86}
\hat{\rho}+\frac{\hat{E}^{2}}{8\pi}-|\hat{p}_{t}+\frac{\hat{E}^{2}}{8\pi}|\geq
0.
\end{eqnarray}
Figures \textbf{18-23} indicate that these inequalities hold for
each $x\in[0,1]$.

\subsection{The van der Waals (VDW) EoS}

Lobo \cite{14} introduced this type of bounded source for the sake
of cosmology. Here we express this approach with VDW EoS given by
\begin{equation}\label{87}
p_{r}=\frac{\omega_{1}\rho}{1-\omega_{3}\rho}-\omega_{2}\rho^{2}.
\end{equation}
This is used to describe dark matter and dark energy as a single
fluid. It is assumed that the interior and exterior metrics are
joined and $E(0)=0=p_r(a)$. Further, $p_r$ is found at the center
and boundary of the charged cylinder. Using Eqs.(\ref{87}) and
(\ref{32}), it follows that
\begin{equation}\label{88}
\omega_{2}\omega_{3}\rho^{3}-(\omega_{2}+\omega_{3})\rho^{2}
+[1+\omega_{1}+\omega_{3}l(r)]\rho-l(r)=0,
\end{equation}
where $\omega_{1}$ and $\omega_{2}$ are functions of
$\rho(0),~\rho(a),~p_{r}(0)$ and $\omega_{3}$. We do not expect any
interesting consequences from this equation, hence we leave it here.

\subsection{Equilibrium Condition}

Now we discuss the variation of the net charge corresponding to
different forces compatible with equilibrium configuration for our
models. In particular, when pressure gradients tend to zero and the
charged fluid is more diluted, then what is the behavior of
gravitational and other forces. Using Tolman-Oppenheimer-Volkov
equation, we obtain
\begin{equation}\label{89}
(\rho+p_{r})(e^{2\nu}(p_{r}-E^{2})-e^{4\nu-2\mu}\frac{\mu'}{r})
-\frac{dp_{r}}{dr}+\frac{\sigma
qe^{\mu-\nu}}{r^{2}}+r(e^{2\mu+2\nu})(p_{t}-p_{r})=0.
\end{equation}
This provides the equilibrium condition for the charged fluid
elements subject to different forces. Here $q=q(r)$ as given in
Eq.(\ref{80}). In adimensional version, we can write
\begin{equation}\label{90}
\hat{F}_{1}+\hat{F}_{2}+\hat{F}_{3}+\hat{F}_{4}=0,
\end{equation}
where
\begin{eqnarray}\label{91}
\hat{F}_{1}&=&(\hat{\rho}+\hat{p}_{r})(e^{\hat{2B}x^{2}+2C}
(\hat{p}_{r}-\hat{E}^{2})-2\hat{A}xe^{4\hat{B}x^{2}-2\hat{A}x^{2}+4C}),\\\label{92}
\hat{F}_{2}&=&-\frac{d\hat{p}_{r}}{dx},\\\label{93}
\hat{F}_{3}&=&\hat{\sigma}\hat{E}e^{x^{2}(\hat{B}-\hat{A}+C)},\\\label{94}
\hat{F}_{4}&=&xe^{2x^{2}(\hat{B}-\hat{A}+2C)}(\hat{p}_{t}-\hat{p}_{r}).
\end{eqnarray}
The graphs of these forces with linear EoS at $\chi=0$ and
$\chi=0.56$ are shown in Figures \textbf{24} and \textbf{25}
respectively which indicate that the charge has a negligible effect
on these forces, hence we obtain a static equilibrium. For
$\chi=0,~\hat{F_{4}}$ point outwards at every $x\in(0,1]$ and
$\hat{F_{2}}$ is along the $x$-axis. The electric force
$\hat{F_{3}}$ acting on the fluid elements with unbounded
$\hat{\sigma}$ located at $x=1$ is infinite. This is the weakest
force because it changes sign at $x\approx 0.38$ and also the force
$\hat{F}_{1}$ changes sign at $x \approx 0.9$ which is due to
gravity. When $\chi=0.56$, the electric force is still unbounded and
infinite. This unboundedness of the force $\hat{F}_{3}$ and sign
inversion of the force $\hat{F}_{1}$ is essential for the
configuration of our static, charged anisotropic model with linear
EoS.

\section{Outlook}

The main purpose of this paper is to investigate the solutions of
the coupled Einstein-Maxwell field equations for the static
cylindrically symmetric spacetime. For this purpose, we have used
charged anisotropic fluid with EoS in the light of Victor et al.
\cite{7} procedure developed for static spherically symmetric
spacetime. In particular, the linear, nonlinear and Chaplygin EoS
have been used. It is mentioned here that the linear EoS
corresponds to the electrically charged isotropic strange quark
stars on the basis of MIT bag model \cite{22}. Our model with
linear EoS also corresponds to the model by Victor et al. \cite{7}
for the charged anisotropic spherically symmetric fluid with the
same EoS. We know that the nonlinear and Chaplygin EoS are used to
describe non-static neutral gravitational isotropic fluids. Here
we are taking static charged anisotropic fluid as we would like to
explore the interior regions and the fluid vacuum interfaces of
the charged anisotropic cylindrically symmetric stars with these
EoS.

We have used the assumptions of Karori and Barua to explore the
charged anisotropic static cylinder. Our models with nonlinear and
Chaplygin EoS correspond to the dark matter and dark energy with
constant matter densities and pressures \cite{23}. Delgaty and Lake
\cite{26} proposed some physical conditions acceptable for perfect
fluids, i.e., regularity of the charge at the origin, positive
matter density and pressure, decreasing matter density and pressure
with increasing $r$, causal sound propagation and smooth matching of
internal and external metrics at boundary of the source. Burke and
Hobill \cite{27} added one more condition that sound velocity must
be monotonically decreasing with increasing $r$ which was imposed on
spherical perfect fluid \cite{28,29}. We have found that our
cylindrical models satisfy most of the above physical conditions. In
our case, conflict may arise for increasing tangential sound
velocity, constant radial velocity and negative tangential pressure
for higher values of $\chi$.

Finally, we would like point out that we have also explained
charged anisotropic static cylindrically symmetric models in our
recent paper \cite{12a} by using Thirukkanesh and Maharaj
approach. In that paper, we have found that the charge
distribution as well as $E$ become singular at $r=0$. However,
here we have found non-singular behavior of these quantities.

\begin{figure}
\center\epsfig{file=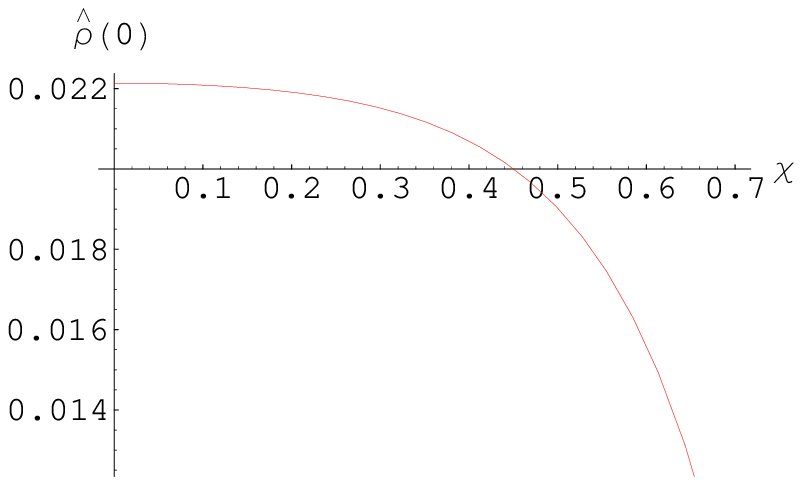, width=0.5\linewidth} \caption{Variation
of central density $\hat{\rho}(0)$ with $\chi$.}
\center\epsfig{file=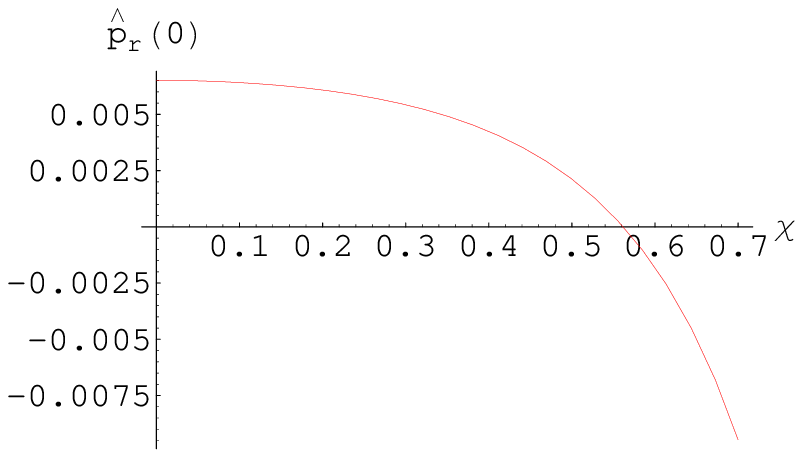, width=0.5\linewidth} \caption{Variation
of central pressure $\hat{p}_{r}(0)$ with $\chi$.}
\center\epsfig{file=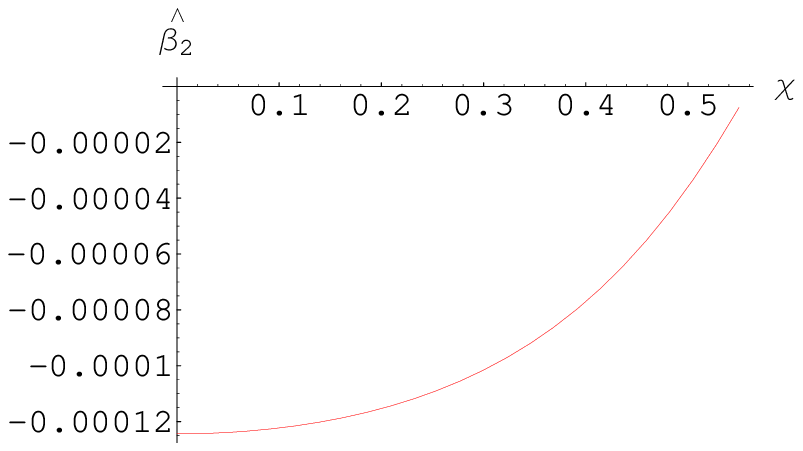, width=0.5\linewidth} \caption{Variation
of parameter $\hat{\beta}_{2}$ with $\chi$.}
\end{figure}
\begin{figure}
\center\epsfig{file=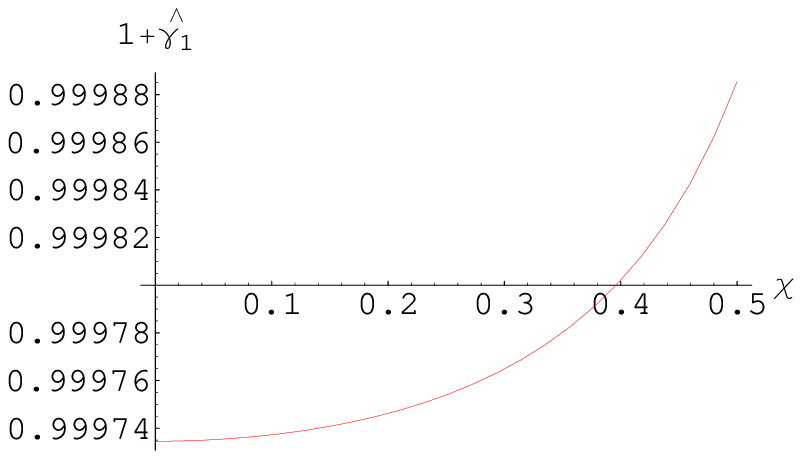, width=0.5\linewidth}
\caption{Variation of parameter $\hat{\gamma}_{1}$ with $\chi$.}
\center\epsfig{file=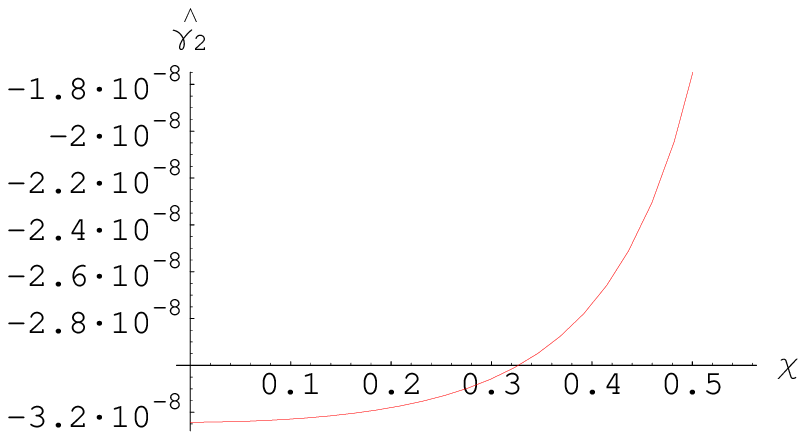, width=0.5\linewidth}
\caption{Variation of parameter $\hat{\gamma}_{2}$ with $\chi$.}
\center\epsfig{file=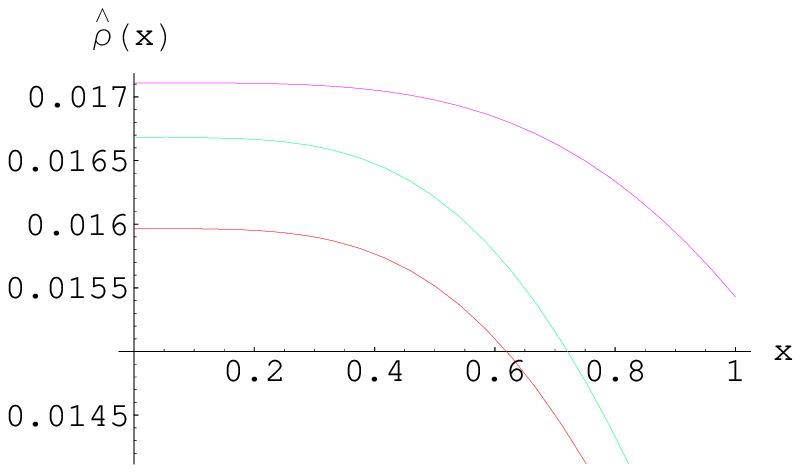, width=0.5\linewidth}
\caption{Variation of matter density $\hat{\rho}(x)$ with $x$ for
different values of $\chi$. Red, green and purple curves
correspond to $\chi=0, 0.3, 0.56$ respectively. }
\end{figure}
\begin{figure}
\center\epsfig{file=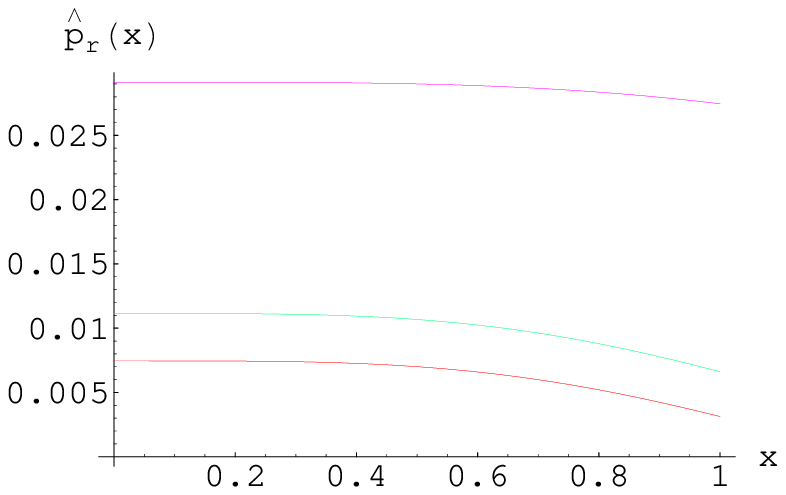, width=0.5\linewidth}
\caption{Variation of radial pressure $\hat{p}_{r}(x)$ with $x$
for different values of $\chi$. Red, green and purple curves
correspond to $\chi=0, 0.3, 0.56$ respectively.}
\center\epsfig{file=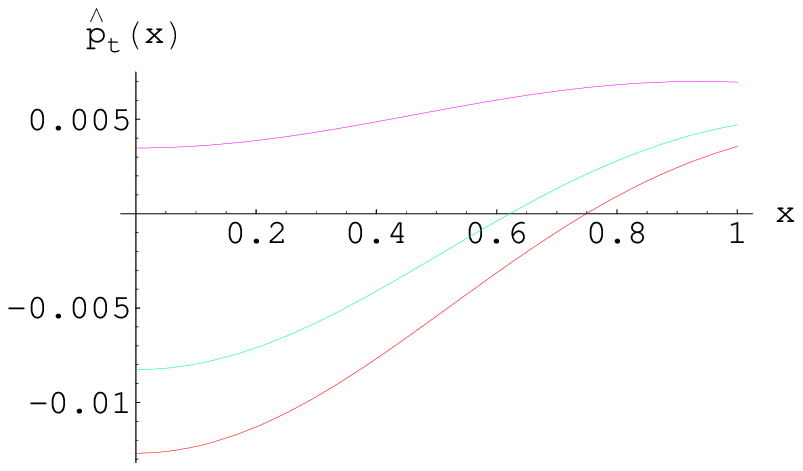, width=0.5\linewidth}
\caption{Variation of tangential pressure $\hat{p}_{t}(x)$ with
$x$ for different values of $\chi$. Red, green and purple curves
correspond to $\chi=0, 0.3, 0.56$ respectively.}
\center\epsfig{file=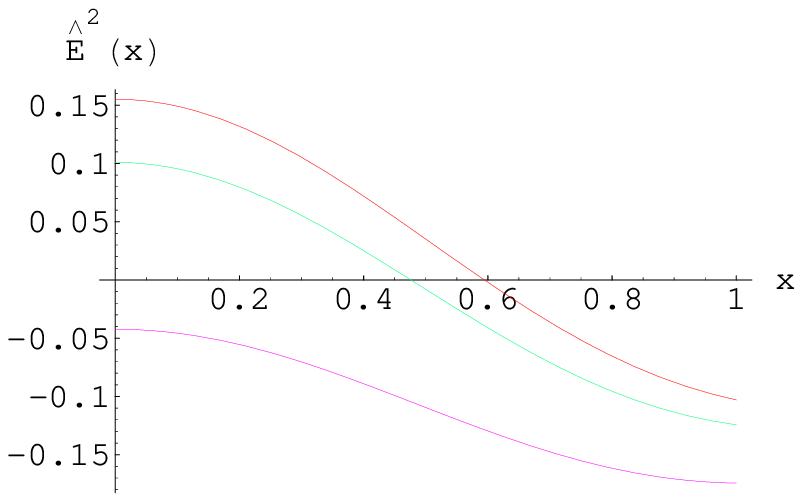, width=0.5\linewidth}
\caption{Variation of Electric field intensity $\hat{E}(x)$ with
$x$ for different values of $\chi$. Red, green and purple lines
correspond to $\chi=0, 0.3, 0.56$ respectively.}
\end{figure}
\begin{figure}
\center\epsfig{file=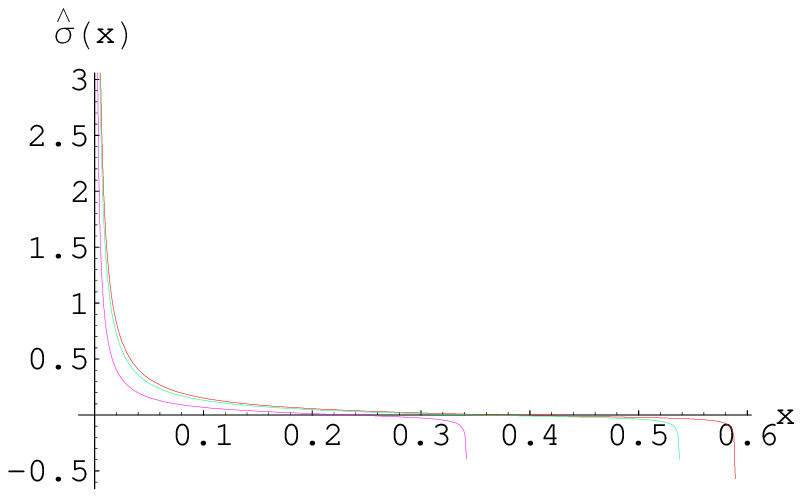, width=0.5\linewidth} \caption{The
charge density $\hat{\sigma}(x)$ for different values of $\chi$.
Red, green and purple curves correspond to $\chi=0, 0.3, 0.56$
respectively.} \center\epsfig{file=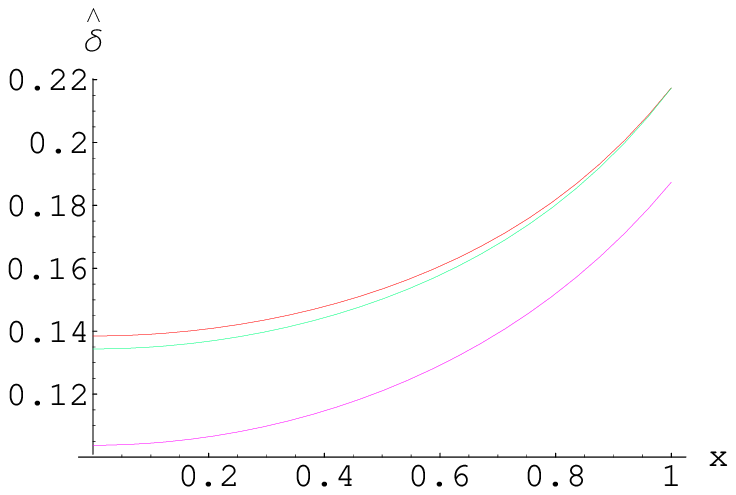, width=0.5\linewidth}
\caption{Measure of anisotropy $\hat{\delta}$ for the first model
with $x$ for different values of $\chi$. Red, green and purple
curves correspond to $\chi=0, 0.3, 0.56$ respectively. }
\center\epsfig{file=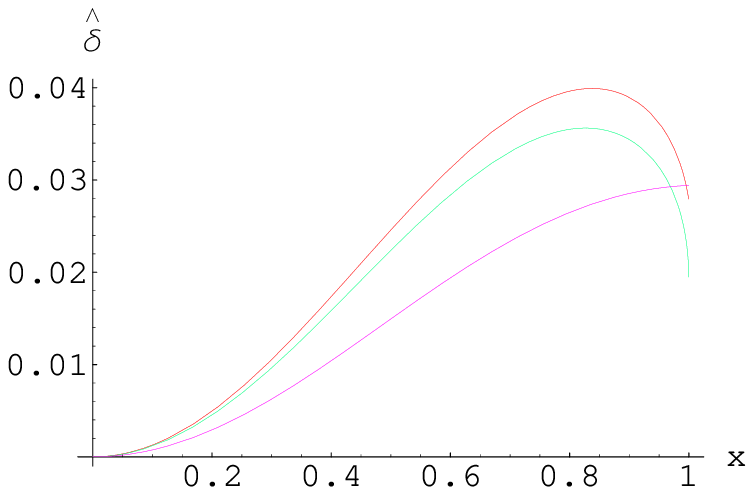, width=0.5\linewidth} \caption{Measure
of anisotropy $\hat{\delta}$ for the second model with $x$ for
different values of $\chi$. Red, green and purple lines correspond
to $\chi=0, 0.3, 0.56$ respectively. }
\end{figure}
\begin{figure}
\center\epsfig{file=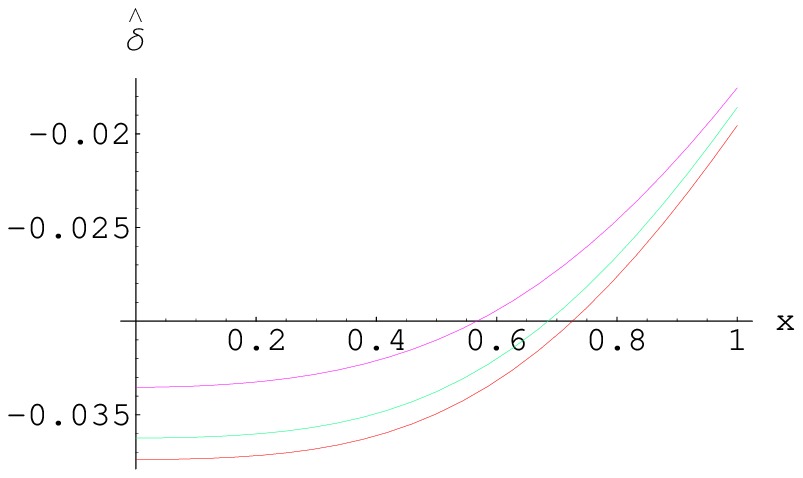, width=0.5\linewidth} \caption{Measure
of anisotropy $\hat{\delta}$ for the third model with $x$ for
different values of $\chi$. Red, green and purple lines correspond
to $\chi=0, 0.3, 0.56$ respectively.} \center\epsfig{file=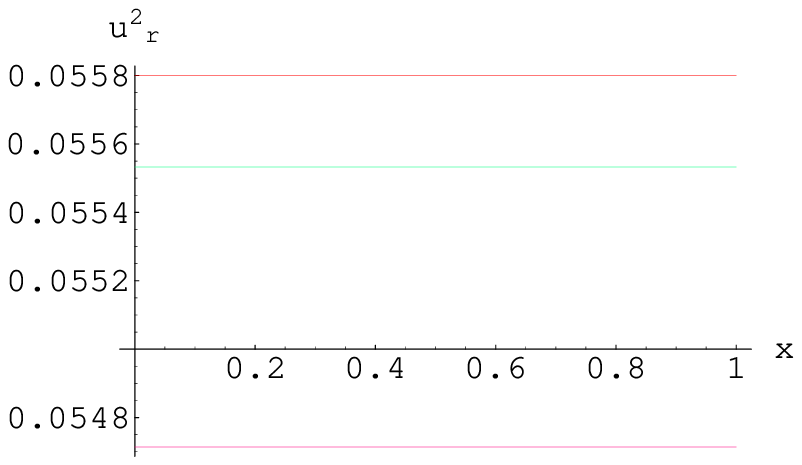,
width=0.5\linewidth} \caption{Variation of radial sound velocity
$\hat{u}_{r}(x)$ with $x$ for different values of $\chi$. Red,
green and purple lines correspond to $\chi=0, 0.3, 0.56$
respectively.} \center\epsfig{file=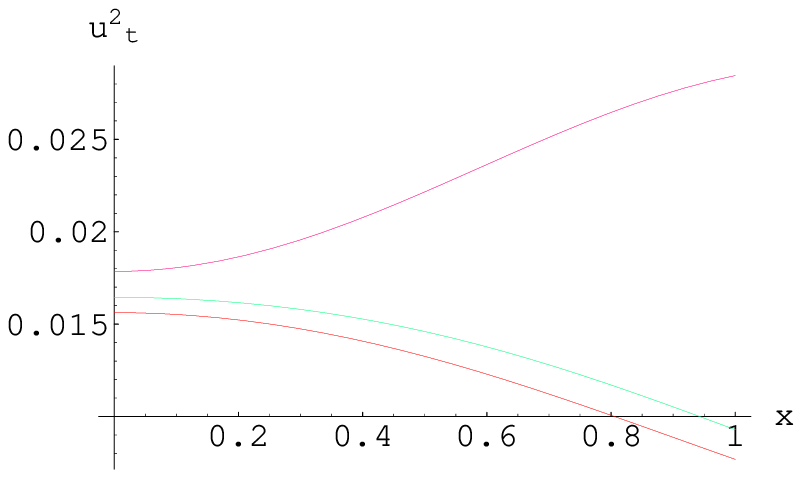, width=0.5\linewidth}
\caption{Variation of tangential sound velocity $\hat{u}_{t}(x)$
with $x$ for different values of $\chi$. Red, green and purple
curves correspond to $\chi=0, 0.3, 0.56$ respectively.}
\end{figure}
\begin{figure}
\center\epsfig{file=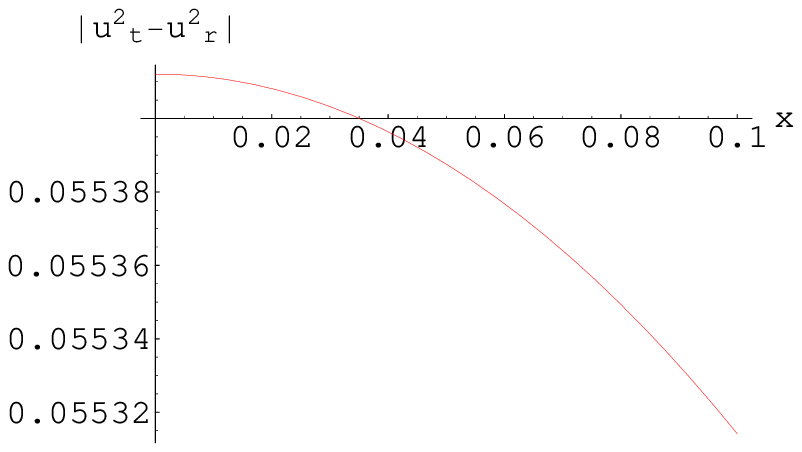, width=0.5\linewidth}
\caption{$|u_{t}^{2}-u_{r}^{2}|\leq 1$} \center\epsfig{file=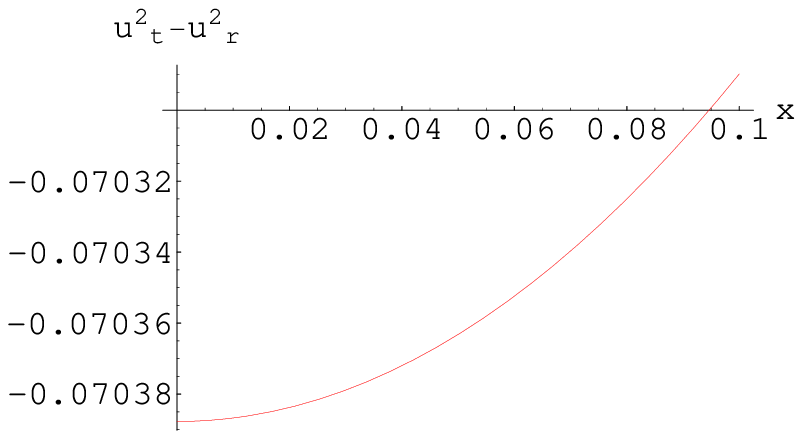,
width=0.5\linewidth} \caption{$-1<u_{t}^{2}-u_{r}^{2}\leq 0$}
\end{figure}
\begin{figure}
\center\epsfig{file=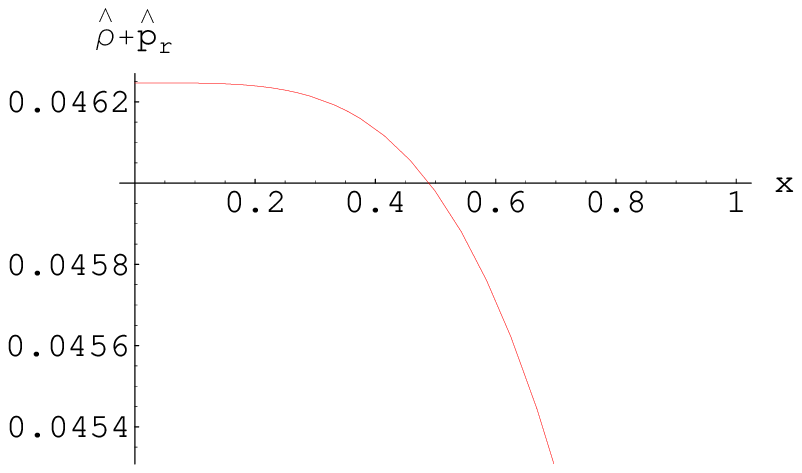, width=0.5\linewidth} \caption{Weak
Energy Condition 1} \center\epsfig{file=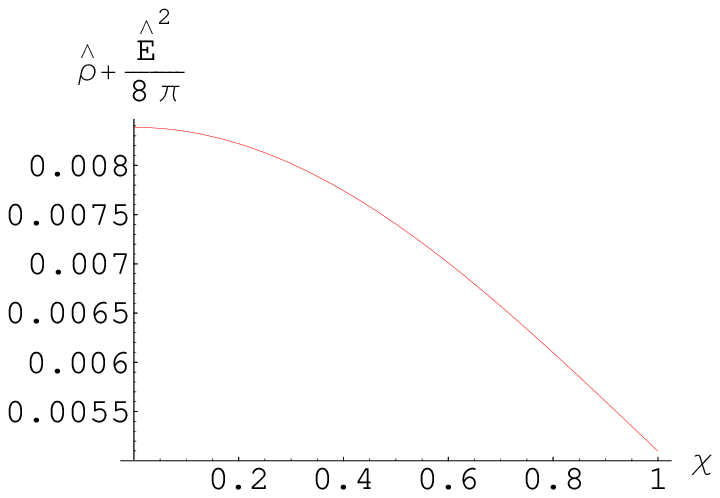,
width=0.5\linewidth} \caption{Weak Energy Condition 2}
\center\epsfig{file=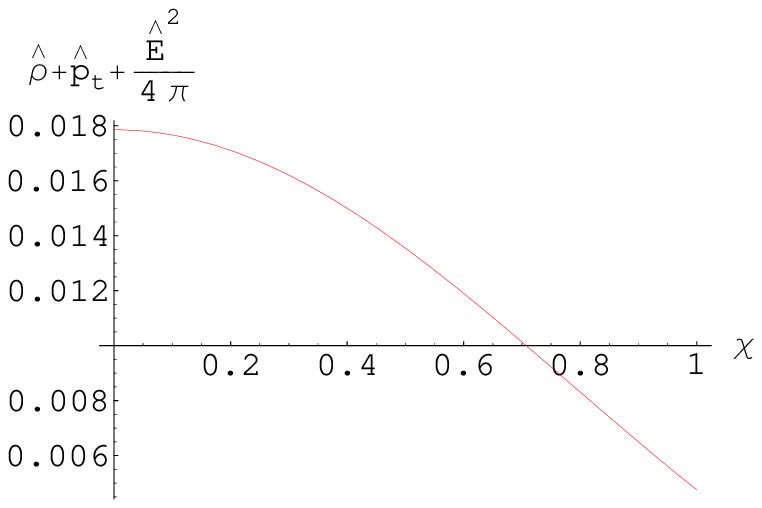, width=0.5\linewidth} \caption{Strong
Energy Condition 1}
\end{figure}
\begin{figure}
\center\epsfig{file=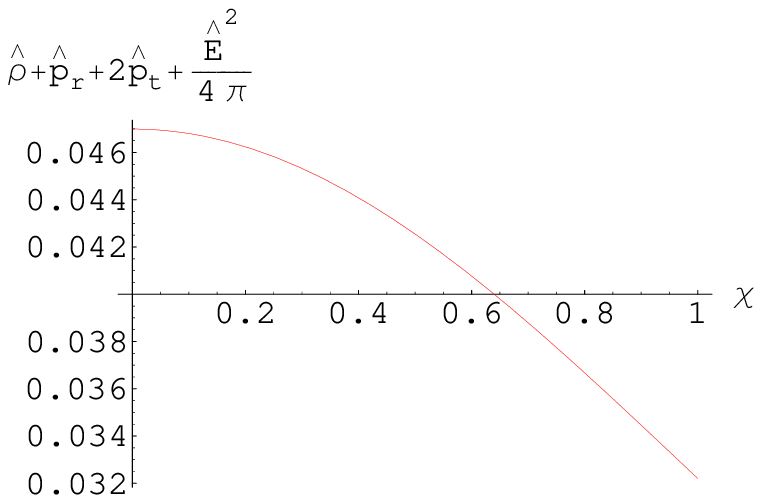, width=0.5\linewidth} \caption{Strong
Energy Condition 2} \center\epsfig{file=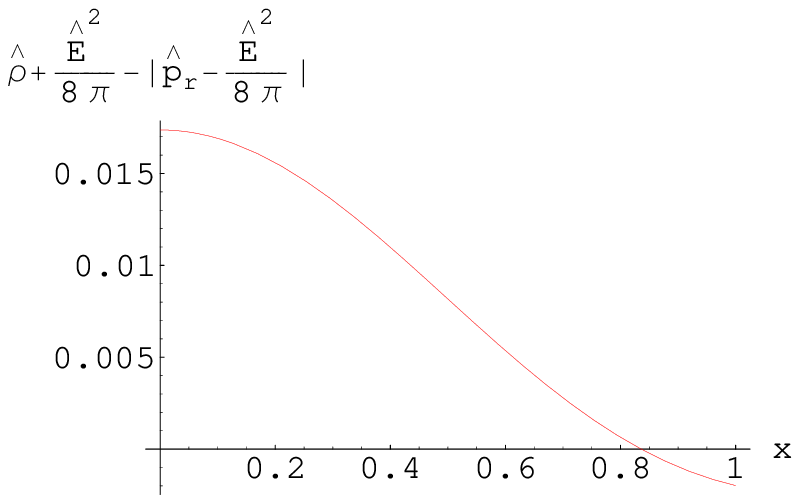,
width=0.5\linewidth} \caption{Dominant Energy Condition 1}
\center\epsfig{file=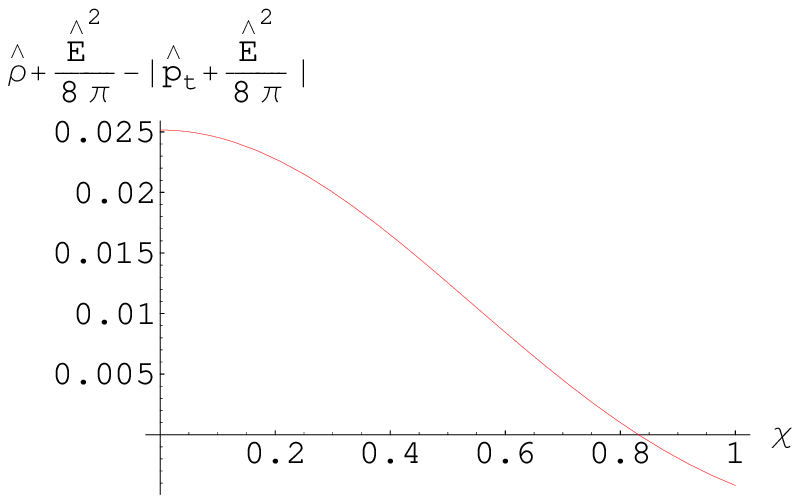, width=0.5\linewidth} \caption{Dominant
Energy Condition 2}
\end{figure}
\begin{figure}
\center\epsfig{file=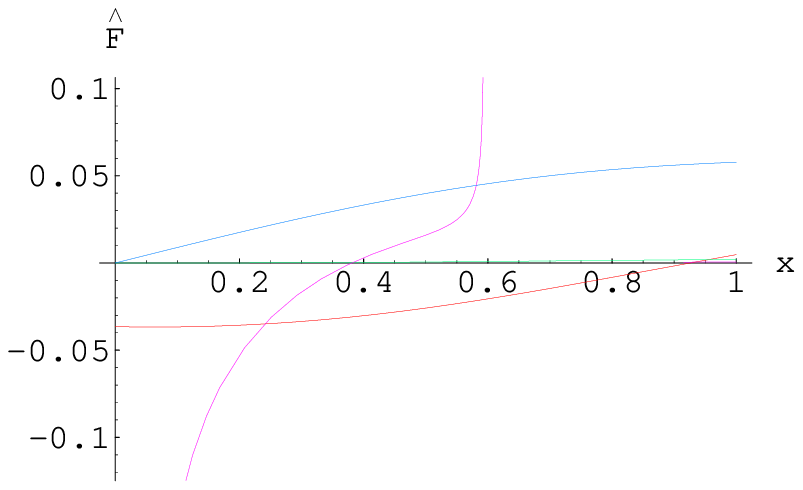, width=0.5\linewidth} \caption{Four
different forces $\hat{F}_{1}(x)$, $\hat{F}_{2}(x)$,
$\hat{F}_{3}(x)$ and $\hat{F}_{4}(x)$ corresponding to red, green,
purple and blue curves respectively are acting on the fluid
elements in static equilibrium for $\chi=0$.}
\center\epsfig{file=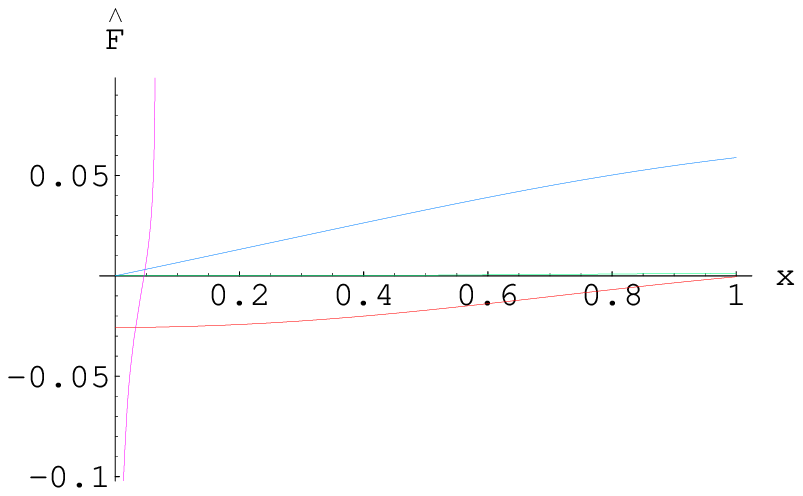, width=0.5\linewidth} \caption{Four
different forces $\hat{F}_{1}(x)$, $\hat{F}_{2}(x)$,
$\hat{F}_{3}(x)$ and $\hat{F}_{4}(x)$ corresponding to red, green,
purple and blue curves respectively are acting on the fluid
elements in static equilibrium for $\chi=0.56$.}
\end{figure}

\end{document}